\documentclass[aps,preprint,a4paper,showpacs,amsmath,amssymb,floatfix]{revtex4}
\input psfig.sty
\newcommand{\Op}[1]{{\boldsymbol{\mathrm{\hat{#1}}}}}
\begin{document}
\draft

\title{Characteristics of the Limit Cycle of a Reciprocating Quantum Heat Engine.}

\author{Tova Feldmann and Ronnie Kosloff}

\address{
Department of Physical Chemistry
the Hebrew University, Jerusalem 91904, Israel\\
}

\begin{abstract}
When a  reciprocating heat engine is started it eventually settles
to a stable mode of operation. The approach of a first principle
quantum heat engine toward this stable limit cycle is studied. The
engine is based on a working medium consisting of an ensemble of
quantum systems composed of two coupled spins. A four stroke cycle
of operation is studied, with two {\em isochore} branches where
heat is transferred from the hot/cold baths and two {\em adiabats}
where work is exchanged. The dynamics is generated by a completely
positive map. It has been shown that the performance of this model
resembles an engine with intrinsic friction. The quantum
conditional entropy is employed to prove the monotonic approach to
a limit cycle. Other convex measures, such as the quantum distance
display the same monotonic approach. The equations of motion of
the engine are solved for the different branches and are combined
to a global propagator that relates the state of the engine in the
beginning of the cycle to the state after one period of operation
of the cycle. The eigenvalues of the propagator define the rate of
relaxation toward the limit cycle. A longitudinal and transverse
mode of approach to the limit cycle is identified. The entropy
balance is used to explore the necessary conditions which lead to
a stable limit cycle. The phenomena of friction can be identified
with a zero change in the von Neumann entropy of the working
medium.
\end{abstract}

\maketitle
\section{Introduction}
\label{sec:introduction}

On starting up a reciprocating four stroke engine, after a few
cycles, the engine settles to a limiting smooth cycle of
operation. The present theoretical analysis is devoted to the
characterization of the transition period from the state when the
engine is started up to the sequence of states characterizing the
periodic steady state termed the limit cycle. The process has
similarities to the approach to thermodynamical equilibrium of an
initially displaced state. This relaxation to equilibrium is
accompanied by entropy production signifying the irreversible
character of the process. Entropy is produced in the approach to
the limit cycle but unlike an equilibrium state entropy continues
to be produced also when the limit cycle is reached.

The approach to a limit cycle is based on the concept of a basin
of attraction where the limit cycle is located at its minimum. The
basin of attraction is set by external and internal constraints.
Dissipative forces cause the system to settle down to the minimum
of such a basin. A first principle  study requires to determine
the equations of motion governing the dynamics of the engine. For
this task in the study, the framework of open quantum systems is
employed \cite{alicki87,lindblad76}. The key point is that the
dynamics of the engine is governed by a completely positive map
\cite{kraus71}. Then the limit cycle becomes a fixed point of this
map. In order to determine if the  approach to the limit cycle is
monotonic, a measure of distance between the actual state of the
engine and the final limiting cycle has to be defined. Such a
measure of distance between two quantum states is not obvious due
to the possibility that the two states do not commute. In analogy
to linear response theory it is expected that close enough to the
limit cycle all distance measures should show the same relaxation
rate toward the target limit cycle. This prediction is consistent
with the results of the present study. Nevertheless at large
distance from the limit cycle only the quantum measures show a
monotonic approach to it.

The present paper is a continuation of a series of studies on a
four stroke quantum engine
\cite{geva0,feldmann96,feldmann00,kosloff01,kosloff03}. The models
studied were based on a first principle quantum description of the
dynamics. The previous studies showed that the model engine
displays the irreversible characteristics of common real heat
engines operating in finite time. The performance of the quantum
engine was found to be limited by finite heat transfer. In
addition quantum performance limitations on the adiabatic branches
mimicked very closely macroscopic friction phenomena
\cite{kosloff01}.

The quantum discrete heat engine is composed of a quantum working
fluid, a hot and a cold bath and an external field which can alter
the energy levels of the working medium. The control parameters
are the time allocations on the different branches, the total
cycle time and the extreme values of the external field. All four
branches are described by quantum equations of motion. The
thermodynamical consequences can therefore be derived from first
principles. A minimum set of three thermodynamical observables was
found which were sufficient to characterize the performance of the
engine. With two additional variables, the state of the working
fluid could also be characterized \cite{kosloff03}. Knowledge of
the state is necessary in order to evaluate the entropy and the
internal temperature, variables which are necessary to establish a
thermodynamic perspective.

The intuitive notion is that the limit cycle is characterized by
the external constraints and internal properties of the engine.
The following questions arise naturally:
\begin{itemize}
\item{How do the control parameters characterize the approach to
the limit cycle?} \item{Can conditions be found for the
non-existence of a limit cycle?}\item{What are the irreversible
properties of the limit cycle?}
\end{itemize}
The present paper is devoted to the study of these issues in the context of
quantum thermodynamics.

\section{Quantum Thermodynamical Observables and their Dynamics}
\label{sec:sumquantherob}

In the field of quantum thermodynamics, thermodynamical variables
are associated with quantum mechanical observables. An observable
$\langle {\bf \hat A} \rangle $, is defined as the following
scalar product between the operator ${\bf \hat A}$ and the density
operator ${\Op \rho}$:
\begin{equation}
\langle {\bf \hat A} \rangle ~=~ \left({\bf \hat A}\cdot { \Op \rho} \right)
~=~ tr \{ {\bf \hat A}^{\dag} ~ {\Op  \rho} \}~~~.
\label{eq:sclar}
\end{equation}
The dynamics of the quantum thermodynamical observables are
described by completely positive maps within the formulation of quantum open systems
\cite{lindblad76,alicki87}. The dynamics is
generated by the Liouville super operator, which in the Heisenberg
picture becomes:
\begin{equation}
\dot {\bf \hat A}~~=~~ {\cal L}^* ( {\bf \hat A})
~+~ \frac{\partial {\bf \hat A}}{\partial t}~~~.
\label{eq:heisenberg}
\end{equation}
where ${\cal L}$ is a generator of a completely positive map:
${\cal T}(t) = e^{{\cal L} t}$. The generator ${\cal L}$ can be
decomposed to the unitary and dissipative contributions ${\cal
L}^*={\cal L}_H^*+{\cal L}_D^*$. The second term in Eq.
(\ref{eq:heisenberg}) $\frac{\partial {\bf \hat A}}{\partial t}$,
addresses a possible explicit time dependence of the operator.

The thermodynamical construction follows Gibbs by seeking a
minimum set of variables associated with the quantum orthogonal
observables $\{ \Op B_k \}$. This set should be sufficient to
completely determine the state of the system $\Op\rho$. In
addition the set should be closed for the dynamics i.e for the
operation of ${\cal L}^*$. Any cycle of a heat engine can be
decomposed into a sequence of four completely positive maps
defining the different branches. Eventually this sequence closes
upon itself. A thermodynamical description therefore means that
the set of variables should be closed for the dynamics during all
branches of the operation. In equilibrium statistical mechanics
the energy of a subsystem is sufficient to determine its state. In
the present non-equilibrium example, additional variables $\langle
{\Op B}_k \rangle $ are required to define the state of the
working fluid. The set of time dependent expectation values $\vec
{\bf b}(t)$ are used to reconstruct the density operator:
\begin{equation}
{\Op \rho}~~=~~\frac{1}{N} {\Op I} + \sum_{\bf k} b_{\bf k} {\Op
B_k}~~~, \label{eq:dens}
\end{equation}
where the expansion coefficients become $b_{\bf k} =\langle {\Op
B}_k \rangle $, $\left( {\Op B}_k \cdot {\Op B}_j \right) = tr \{
\Op B_k^{\dagger} \Op B_j \} = \delta_{kj}$, $tr \{\Op B_k\}=0$,
and N is the size of the Hilbert space.

\subsection{Quantum entropy}

Thermodynamic measures require the knowledge of the state of the
system $\Op \rho$. Entropy, the most common measure, is associated
with  the lack of knowledge or dispersion of the system
\cite{ruskai02,Vedral02}. The entropy associated with a
measurement of an observable $\langle {\bf \hat A} \rangle $ with
$N$ possible outcomes becomes:
\begin{equation}
S_{\Op A} ~~=~~ - \sum_j^N p_j \log p_j~~,
\label{eq:entropya}
\end{equation}
where $p_j = tr \{ \Op P_j \Op \rho \}$ and $\Op P_j$ is the $j$
projection operator of the operator $\Op A = \sum_j^N \alpha_j \Op
P_j $, and where the spectral decomposition ($\Op A|\phi_j\rangle
= \alpha_j|\phi_j\rangle$, $ \Op P_j=| \phi_j\rangle \langle
\phi_j |$) was utilized. Looking for the observable which complete
measurement maximizes the information on the state, is equivalent
to minimizing the entropy with respect to all possible
observables. This process leads to the von Neumann entropy
\begin{equation}
{\cal S} ~~=~~ - tr \{ \Op \rho \log \Op \rho \}~~,
\label{eq:entropyv}
\end{equation}
and the optimum operator that minimizes dispersion commutes with
the state $\Op \rho$.

The distance from a reference state $\Op \rho_{ref}$ is a key
component in the study of the approach to the equilibrium or to
the  steady state. The conditional entropy is associated with the
lack of information on the state $\Op \rho$ subject to the
knowledge of a reference state $\Op \rho_{ref}$. The conditional
entropy  associated with a measurement of a particular observable
becomes:
\begin{equation}
{\cal S}_{\Op A}(\Op \rho | \Op \rho_{ref} ) ~~=~~ - \sum_j p_j \log
\frac {p_j}{q_j} ~~,
\label{eq:relentropya}
\end{equation}
where $q_j= tr \{ \Op P_j \Op \rho_{ref} \}$.
The conditional entropy is bound from above and positive:
\begin{equation}
{\cal S}_{\Op A}(\Op \rho) \geq S_{\Op A}(\Op \rho | \Op \rho_{ref} )~~.
\geq 0 \label{eq:spositive}
\end{equation}
The value zero is reached only when $\Op \rho=\Op \rho_{ref}$.

Maximizing Eq. (\ref{eq:relentropya}) with respect to the
operator $\Op A$, leads to an entropy measure which depends only
on the two states:
\begin{equation}
{\cal S}(\Op \rho | \Op \rho_{ref} ) ~~=~~ - tr \{ \Op \rho ( \log \Op
\rho- \log \Op \rho_{ref} ) \} ~~,
\label{eq:relentropy}
\end{equation}
$S(\Op \rho | \Op \rho_{ref} )=0$ when the two states become
indistinguishable.

\subsection{Conditions for the monotonic approach to the limit cycle}
\label{subsec:mono}

Lindblad \cite{lindblad75} has proven that the conditional entropy
decreases if a completely positive map is applied to both the
state $\Op \rho$ and the reference state $\Op \rho_{ref}$:
\begin{equation}
{\cal S}( \Op \rho |  \Op \rho_{ref} ) ~~\geq~~ {\cal S} ({\cal T} \Op \rho |
{\cal T} \Op \rho_{ref} ) ~~,
\label{eq:inelentropy}
\end{equation}
where ${\cal T}$ is a completely positive map. An interpretation
of Eq. (\ref{eq:inelentropy}) is that a completely positive map
reduces the distinguishability between two states. This observation
has been employed to prove the monotonic approach to equilibrium,
provided that the reference state $\Op \rho_{ref}$ is the {\bf
only} invariant of the mapping $\cal T$ i.e. ${\cal T}\Op
\rho_{ref} = \Op \rho_{ref}$ \cite{frigerio77,frigerio78}.

The same reasoning  can prove the monotonic approach to the limit
cycle. The mapping imposed by the cycle of operation of a heat
engine is a product of the individual evolution steps along the
branches composing the cycle of operation (Cf. \ref{subs:globalp}
). Each one of these evolution steps is a completely positive map,
so that the total evolution ${\cal U}_{cyc}$ that represents one
cycle of operation, is also a completely positive map. If then a
state ${\Op \rho}_{lc}$ is found that is a single invariant of
${\cal U}_{cyc}$ i.e. ${\cal U}_{cyc} \Op \rho_{lc} = \Op
\rho_{lc}$ then any initial state ${\Op \rho}_{init}$  will
monotonically approach to the limit cycle.  Based on Eq.
(\ref{eq:inelentropy}) a monotonic decreasing series bound from
below converges to a limit.
\begin{equation}
S( \Op \rho_{init} |  \Op \rho_{lc} ) \geq
S({\cal U}_{cyc} \Op \rho_{init} |  \Op \rho_{lc} ) \geq
S({\cal U}_{cyc}^n \Op \rho_{init} |  \Op \rho_{lc} ) ~~\geq ~~0~~,
\label{eq:limit1}
\end{equation}
where $ {\cal U}_{cyc}^n$ represents a sequential mapping of the cycle $n$ times.

The largest eigenvalue of ${\cal U}_{cyc}$ with a value of one is
associated with the invariant limit cycle state
${\cal U}_{cyc}{\Op \rho}_{lc}={\bf 1} \Op \rho_{lc}$, 
the fixed point of ${\cal U}_{cyc}$. The other
eigenvalues determine the rate of  approach to the limit cycle (Cf. Sec. \ref{susec:approach})

The conditional entropy has been criticized as a measure of
distance since it is not symmetric in $\Op \rho$ and $\Op
\rho_{ref}$ and therefore does not form a metric. For this reason
other measures have been defined.

\subsection{Thermodynamic Quantum Distance}
\label{subsec:distance}

The concept of statistical distance between different pure quantum
systems was introduced by W. K. Wootters  \cite{wootters81}, who
followed R. A. Fisher's  \cite{fischer22} idea to measure distance
in probability space. In \cite{brancav94} the concept of
distinguishability for neighboring mixed quantum states is
described. H$\ddot{u}$bner \cite{hubner92}   computed explicitly
the distance between two-dimensional density operators, and gave a
general formula for the \em N \rm dimensional distance. A detailed
and clear review on the subject has been presented by Di{\'o}si and
Salamon \cite{diosal99}.

\subsubsection{Wootters Distance}
\label{subsubsec:wdist}

Statistical distance is associated with the size of the statistical
fluctuations occurring in a measurements that distinguishes one state
from another. Two outcomes are distinguishable in a given number
of trials, provided that the difference in actual probabilities is
larger than the size of typical fluctuation. The maximal number
of distinguishable states that can be found between two
probability distributions has been suggested by Wootters
\cite{wootters81} to define the distance between these two states.

Consider two probability distributions $\bf p$ and $\bf q$ obtained
from the same complete measurement of two quantum states
$\Op \rho$ and $\Op \rho_{ref}$ (Cf. Eq. (\ref{eq:relentropya})).
In order to define a distance between between $\bf p$ and $\bf q$
a continuous curve is sought connecting the two distributions.
Taking advantage of normalization and that probabilities are positive,
a change of variable is used: $x_j=\sqrt p_j$ and $y_j=\sqrt q_j$.
The new variables allow a geometric interpretation, they define
points on an $N$ dimensional unit sphere, since $\sum_j^N x_j^2=\sum_j^N y_j^2=1$.
The statistical distance between $\bf p$ and $\bf q$ then becomes
the shortest distance on the surface of this unit sphere between the points
defined by the vectors $\bf x$ and $\bf y$. This shortest distance is equal to the angle
between the unit vectors $\bf x$ and $\bf y$, given by:
\begin{eqnarray}
{\cal D}_{\Op A}({\bf p},{\bf q} )~=~\arccos \left( \sum_{j=1}^{N}x_j y_j \right)~=~
~~\arccos \left( \sum_{j=1}^{N}\sqrt{p_{j}} \sqrt{q_{j}} \right)~~~.
\label{eq:fdistan}
\end{eqnarray}
For quantum systems Eq. (\ref{eq:fdistan}) corresponds to the
statistical distance associated with a measurement of an operator
$\Op A$. For two commuting states the statistical distance becomes
the $\arccos$ of  scalar product of  the square roots of the density operators.
\begin{eqnarray}
{\cal D}(\Op \rho,\Op \rho_{ref})~=~ ~~\arccos \left(
 tr \left \{ \Op \rho^{1/2} \Op \rho_{ref}^{1/2} \right \} \right)~~.
 \label{fdistan2}
\end{eqnarray}
For two non-commuting states, the distance ${\cal D}(\Op \rho,\Op
\rho_{ref})$ has to be redefined as \cite{hubner92}:
\begin{eqnarray}
{\cal D} (\Op \rho,\Op {\rho}_{ref})~=\sqrt{inf (tr\{(\Op W_1- \Op W_2)(\Op W_1-\Op W_2)^*\}} 
~~,
\label{quantdistan}
\end{eqnarray}
where the infimum is taken over all Hilbert-Schmidt  operators describing all
the possible operators which fulfill
\begin{eqnarray}
\Op W_1 \Op W_1^* = \Op \rho, ~~\Op W_2 \Op W_2^* = \Op \rho_{ref}~~,
\label{quantdist1}
\end{eqnarray}
and $\Op W_1^* \Op W_2>0$. This definition of distance
is symmetric in $\Op \rho$, $\Op \rho_{ref}$ and therefore can form a metric
\cite{hubner92,diosal99}:
\begin{eqnarray}
{\cal D}(\Op \rho,\Op \rho_{ref})~=~\sqrt{N \left(1~-~tr \sqrt{
 (\Op \rho)^{\frac{1}{2}} \Op \rho_{ref} (\Op \rho)^{\frac{1}{2}}}\right)}~~,
\label{eq:quantdist2}
\end{eqnarray}
where $N$ is the size of the Hilbert space.

\subsection{Entropy production}

Once the limit cycle is reached any observable is cyclic including
the entropy. This is the result of the fact that the state of the system
is completely determined by a finite number of expectation values which are
cyclic. This means that entropy change in the complete cycle is zero
or the total internal entropy production of the working medium  is zero.

The external entropy production is positive for the limit cycle.
It is a measure of the irreversible dissipation to the hot and cold baths:
\begin{equation}
\Delta { S}_{cyl}^{ext} ~~=~~ -\left( \frac{ {\cal
Q}_h}{T_h}+\frac{ {\cal Q}_c}{T_c} \right) ~~,
\label{eq:entrprod1}
\end{equation}
where ${\cal Q}_{h/c}$ is the heat dissipated to the hot/cold
bath and $T_{h/c}$ is the bath temperature.

\section{The quantum model}
\label{sec:quantumod}

The present study is based on a four stroke quantum heat engine
model corresponding to the Otto cycle. The cycle is composed of
two {\em isochores} where the working medium is in contact with the
hot/cold baths and the external field is constant and two {\em
adiabats} where the external field is varying. The motion is
generated by the Liouville operator $\cal L$ which can be
decomposed to a Hamiltonian part and a dissipative part:
\begin{eqnarray}
{\cal L}^* ~~=~ {\cal L}_H^* +{\cal L}_D^*~~,
\label{eq:liouville}
\end{eqnarray}
where ${\cal L}_H^* {\Op A}=  i [ {\Op H}, {\Op A}]$.
The main feature of the Hamiltonian is that the external control
part does not commute with the inertial internal part.
\begin{equation}
{\Op H} ~~=~~ {\Op H}_{int}+{\Op H}_{ext}(t)~~,
\end{equation}
and $[{\Op H}_{int},{\Op H}_{ext}] ~~ \neq 0$.

The specific choice of working medium is composed of an ensemble of
noninteracting coupled two-spin systems identical to the model studied
in Ref. \cite{kosloff03}.

\subsection{The Hamiltonian}

The single particle Hamiltonian is chosen to be proportional
to the polarization of a two-level-system (TLS):
${\boldsymbol{\mathrm{\hat{\sigma}}}}_z^j$.
The operators ${\boldsymbol{\mathrm{\hat{\sigma}}}}_z,
       {\boldsymbol{\mathrm{\hat{\sigma}}}}_x,
     { \boldsymbol{\mathrm{\hat{\sigma}}}}_y$ are the Pauli matrices.
For this system, the external Hamiltonian will be:
\begin{equation}
{\Op H}_{ext} ~~=~~2^{-3/2} \omega(t)
\left({\boldsymbol{\mathrm{\hat{\sigma}}}}_z^1
\otimes {\bf \hat I^2}
+
{\bf \hat I^1} \otimes {{{{\boldsymbol{ \mathrm {  \sigma}}}}}_z^2}
\right)~\equiv~\omega(t) {\Op B_1}~~,
\label{eq:hext1}
\end{equation}
and the external control field $\omega(t)$ is chosen to
be in the  $z$ direction.
The uncontrolled interaction Hamiltonian is chosen to be
restricted to the coupling
of pairs of spin atoms. Therefore  the working fluid consists
of noninteracting pairs
of TLS's.  For simplicity, a single pair can be considered.
The thermodynamics of $M$ pairs then follows by introducing
a trivial scale factor.
Accordingly let the uncontrolled part be:
\begin{equation}
{\Op H}_{int} ~~=~~2^{-3/2} J \left({ {\boldsymbol{\mathrm{\hat
{\sigma}}}}_x^1} \otimes { {\boldsymbol{\mathrm{\hat{\sigma}}}}_x^2} -
{{\boldsymbol{\mathrm{\hat{\sigma}}}}_y^1}\otimes {\boldsymbol
{\mathrm{\hat{\sigma}}}}_y^2 ~~~.
  \right)~\equiv~J {\Op B_2}~~.
\label{eq:interaction}
\end{equation}
$J$ scales the strength of the interaction. When $J \rightarrow 0$,
the model represents a working medium with noninteracting atoms
\cite{feldmann96}.

The commutation relation:
$ [{\Op B}_{1},{\Op B}_{2}] = \sqrt{2} i {\Op B}_{3}$
leads to the definition of ${\Op B}_{3}$. The analysis shows
\cite{kosloff03}, that the  set of operators
${\Op B}_{1},{\Op B}_{2}, {\Op B}_{3}$ and ${\Op I}$
forms a closed sub-algebra of the total Lie algebra of the combined system.
The Hamiltonian expressed in terms of the operators
${\Op B_1} , {\Op B_2}, {\Op B_3}$ in the polarization representation
becomes:
\begin{eqnarray}
\begin{array}{c}
{ \bf \hat H }
\end{array} ~~=~~
 \begin{array}{c}
\omega(t) {\bf \hat B_{1}}+\rm J {\bf \hat B_{2}}~~.
\end{array}
\label{matHP}
\end{eqnarray}
The eigenvalues of $\Op H$ are $ 0~,~\pm \frac{\Omega}{\sqrt{2}}$ where $\Omega=\sqrt{\omega^2+J^2}$.
The closed Lie algebra of the set $\{ \Op B_k \}$ means that it is also closed for the propagation
generated by the Hamiltonian (\ref{matHP}).

\section{The Cycle of Operation, the Quantum Otto cycle}
\label{sec:cycle}

\begin{figure}[tb]
\vspace{-0.66cm}
\hspace{3.cm}
\psfig{figure=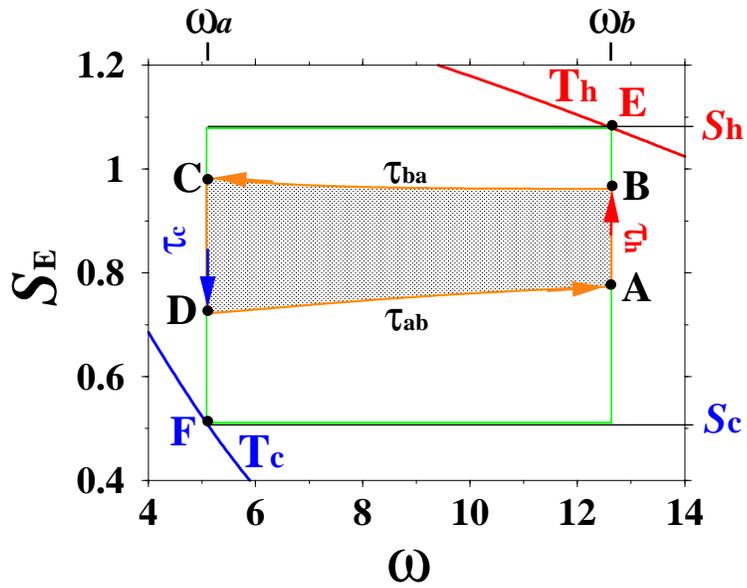,width=0.6\textwidth}
\vspace{0.3cm}
\caption{The cycle of the heat engine in the ($\omega, {\cal S}_E$)
plane. The upper red line indicates the energy entropy of
the working medium in equilibrium with the hot bath at temperature
T$_h$ for different values of the field. The blue line below
indicates the energy entropy in equilibrium with the cold bath at
temperature T$_c$. The cycle in green has an infinite time
allocation on all branches. It reaches the equilibrium point with
the hot bath (point E) and equilibrium point with the cold bath
(point F). The inner cycle ABCD is a typical cycle with the time
allocations: $\tau_h=2.5~ \tau_{ba}=0.01,~\tau_c=3. ~~
\tau_{ab}=0.01   $. The external parameters are: $\omega_a=5.08364
,~\omega_b=12.6355,
~J=2.,~T_h=7.5,~T_c=1.5,~\Gamma_h=~\Gamma_c=0.3423,
~\gamma_h=\gamma_c=0$ } \label{fig:cycle1}
\end{figure}

The operation of the heat engine is determined by the properties
of the working medium and the coupling to the hot and cold baths.
The cycle of operation is defined by the external controls which
include the variation in time of the field with the periodic
property $\omega(t)=\omega(t+\tau)$ where $\tau$ is the total
cycle time synchronized with  the contact times on the different
branches of the cycle. The cycle studied is composed of two
branches   where the working medium is in contact with the
hot/cold baths and the field is constant, termed {\em isochores}.
In addition, there are two branches   where the field $\omega(t)$
varies and the working medium is disconnected from the baths
termed {\em adiabats}. This cycle is a quantum analogue of the
Otto cycle. The four strokes of the cycle with the corresponding
parameters are (Cf. Fig. \ref{fig:cycle1}):
\begin{enumerate}
\item{ {\em Isochore} $A \rightarrow B$: the field is maintained
constant $\omega=\omega_b$ while the working medium
 is in contact with the hot bath of temperature $T_h$ with heat conductance $\Gamma_h$,
and dephasing parameter $\gamma_h$ for a period of $\tau_h$. }
\item{ {\em Adiabat} $B \rightarrow C$: The field  changes linearly
from $\omega_b$ to $\omega_a$ in a time period of $\tau_{ba}$.}
\item{ {\em Isochore} $C \rightarrow D$: the field is maintained
constant $\omega=\omega_a$ the working medium
is in contact with the cold bath of temperature $T_c$
with heat conductance $\Gamma_c$, and dephasing parameter
$\gamma_c$ for a period of $\tau_c$. }
\item{ {\em Adiabat} $D \rightarrow A$: The field changes linearly
from $\omega_a$ to $\omega_b$ in a time period  of $\tau_{ab}$.}
\end{enumerate}

Fig. \ref{fig:cycle1} displays a typical trajectory of the cycle
in the plane defined by the external field and the entropy
($\omega,{\cal S}_E$).

The state $\Op \rho$ (Cf. Eq. (\ref{eq:dens})) of the working
medium is completely reconstructed by the set $\{ {\Op B_k}
\}~,k=1,5$ of five operators \cite{kosloff03}. The set of the
three operators ${\bf \hat B_1}, {\bf \hat B_2}, {\bf \hat B_3}$
is sufficient to describe the energy changes during the cycle of
operation. The map $\cal U$ relates the initial values of these
operators to their final values for each of the engine branches.
Implicitly this map is obtainable by solving a set of coupled
inhomogeneous equations of motion for each branch
\cite{kosloff03}. An explicit description of the map is obtained
by adding the identity operator  $\Op I$ which transforms the
inhomogeneous equations to a closed set of linear coupled $4
\times 4 $ equations. The equation of motion  of the two
additional operators ${\bf \hat B_4}$ and ${\bf \hat B_5}$ form a
linear first order inhomogeneous equation depending on the time
dependence of the closed set ${\bf\hat B_1}, {\bf \hat B_2}, {\bf
\hat B_3} $ and $ {\bf \hat I}$. As a result they form an
additional $2 \times 2 $ block in the map.

\subsection{Propagators on the \em isochores \rm }
\label{subs:propiso}

The dynamical map  ${ \cal U}(\tau_{c/h}) $ on the isochores is generated by both the Hamiltonian
and the dissipative Lindblad generators representing the interaction with the bath.
The $4 \times 4$ block of the map is first solved for the set
${\bf \hat B_1}, {\bf \hat B_2}, {\bf \hat B_3}$ and $\Op I$.
Since $\omega$ is constant on the {\em isochores} a closed form
of the propagator $ { \cal U}(\tau_{c/h})$  is obtained \cite{kosloff03}:
\begin{eqnarray}
\begin{array}{c}
{ \cal U}(\tau_{c/h})=
\end{array}
\left(
\begin{array}{cccc}
 K \frac{X \omega^2+c{J^2}}{{\Omega}^2}&
K \frac{{\omega}J(X-c)}{{\Omega}^2}
 &K \frac{Js}{\Omega}& b_1^{eq}(1-e^{-\Gamma \tau_{c/h}}) \\
K \frac{{\omega}J(X-c)}{{\Omega}^2}&K \frac{X J^2+c{\omega}^2}{{\Omega}^2} &
-K \frac{{\omega}s}{ \Omega}& b_2^{eq}(1-e^{-\Gamma \tau_{c/h}})\\
 -K \frac{Js}{ \Omega} &K \frac{{\omega}s}{\Omega}& K c& 0 \\
0&0&0&1\\
\end{array}
\right)~~~,
\label{PropagTNew}
\end{eqnarray}
where $K=\exp\{-(\Gamma+2 \gamma \Omega^2) \tau_{c/h}\}$,
$X=\exp({2 \gamma \Omega^2 \tau_{c/h}})$, $c=\cos(\sqrt{2}
\Omega \tau_{c/h})$ and $ s=\sin(\sqrt{2} \Omega \tau_{c/h})$,
$b_1^{eq}=-\frac{\omega}{\sqrt{2}\Omega \Gamma}(k\uparrow - k\downarrow)$,
$b_2^{eq}=-\frac{J}{\sqrt{2}\Omega \Gamma}(k\uparrow - k\downarrow)$. Finally
$\gamma$ is the dephasing constant, $ \tau_{c/h}$ is the time
spent on the cold/hot \em isochore \rm, and $\Gamma=k\uparrow +
k\downarrow $ is the heat-conductance to the bath. The corresponding bath
temperature $T_{c/h}$, enters through the detailed balance relation:
$\frac{k \uparrow}{k \downarrow}~~=~~e^{-\frac{ \Omega }{T
\sqrt{2} }}$. The block containing $\Op B_4$ and $\Op B_5$ can now be solved as a
$2 \times 2$ set of coupled inhomogeneous equation of motion \cite{kosloff03}.

\subsection{Propagators on the \em adiabats \rm }
\label{subs:propadiab}

The propagator on the {\em adiabats} is more involved. This is  due to the explicit time
dependence of the Hamiltonian. Using the Lie algebra of the set
of $\{ \Op B_k \}$ operators it is always possible \cite{weinorman63} to describe
the propagator as:
\begin{equation}
\Op A (t) ~~=~~ \Op U \Op A \Op U^\dagger~~,
\label{eq:propt}
\end{equation}
where $\Op U = \exp(i \alpha_1(t) { \Op B_1})\exp(i \alpha_2(t) { \Op B_2})\exp(i \alpha_3(t) { \Op B_3})$
and the coefficients $\alpha_i(t)$ include the effect of time ordering.
The procedure is described in Appendix \ref{ap:ansolprop}.
This approach leads to the explicit result for the propagator of the set
${\bf \hat B_1}, {\bf \hat B_2}, {\bf \hat B_3}, {\bf \hat I}$:
\begin{eqnarray}
 {\cal U}_a(t)      ~~=~~ \left(
\begin{array}{cccc}
c_2c_3 & -s_3c_1+c_3s_2s_1&c_3s_2c_1+s_3s_1&0 \\
c_2s_3   &c_3c_1+s_3s_2s_1& s_3s_2c_1-c_3s_1&0 \\
 -s_2 &   c_2s_1 &c_2c_1 &0\\
0&0&0&1\\
\end{array}
\right)~~~,
\label{propan}
\end{eqnarray}
where:
$s_1=\sin(\alpha_1),~ s_2=\sin(\alpha_2),~ s_3=\sin(\alpha_3)$,~
$c_1=\cos(\alpha_1),~ c_2=\cos(\alpha_2),~ c_3=\cos(\alpha_3) $.
The coefficients $\alpha$ can be integrated either numerically Cf. Eq. (\ref{mateq3})
or closed form solutions are obtained for specific functional forms of $\omega(t)$  \cite{kosloff03}.
The operators $\Op B_4$ and $\Op B_5$ commute with the
Hamiltonian, and therefore  are constant on the \em adiabats \rm.

\subsection{The global propagator }
\label{subs:globalp}

The propagator of the cycle represents the completely positive map of
the initial expectation values to the final ones after the operation of
one cycle. The propagator is then constructed as a sequential product of the individual
propagators on the different branches:
\begin{eqnarray}
{\cal U}_{cyc} = {\cal U}_{ab}~ {\cal U}_{isc}~ {\cal U}_{ba}~
{\cal U}_{ish} ~~. \label{defglobp}
\label{eq:gprop}
\end{eqnarray}
An analytic form has been obtained for the propagators on the {\em isochores} (Eq. (\ref{PropagTNew})).
For the {\em adiabats } the form of Eq. (\ref{propan}) has been used which is parametrically
dependent on the $\alpha$ parameters.

Table I summarizes all the control parameters defining the cycle.
\begin{table}
\caption{Summary of notations}
\begin{tabular}{|ll|}
\hline
$T_c$ & temperature of the cold bath.\\
$T_h$ & temperature of the hot bath.\\
$\omega_a$ &value of the external field at the cold {isochore}.\\
$\omega_b$ &value of the external field at the hot {isochore}.\\
$J$ & internal coupling constant.\\
$\Gamma_c$ & heat transfer coupling constant to the cold bath.\\
$\Gamma_h$ & heat transfer coupling constant to the cold bath.\\
$\gamma_c$ & dephasing constant on the cold bath.\\
$\gamma_c$ & dephasing constant on the hot bath.\\
$\tau_c$ & time allocation on the cold {\em isochore}.\\
$\tau_h$ & time allocation on the hot {\em isochore}.\\
$\tau_{ab}$ & time allocation on the cold-to-hot {\em adiabat}.\\
$\tau_{ba}$ & time allocation on the hot-to-cold {\em adiabat}.\\
\hline
\end{tabular}
\end{table}

The global map enable to solve for the operator expectation values from
their initial values. These expectation values serve to reconstruct
the density operator  (Cf. Eq. (\ref{eq:dens}) ) \cite{kosloff03}:
\begin{eqnarray}
\begin{array}{c}
{\Op \rho_p }
\end{array} ~~=
~~~  \left(
\begin{array}{cccc}
\frac{1 }{ 4}+ \frac{{b_1} }{ \sqrt{2}}+\frac{b_5 }{ 2} & 0 & 0 &
  \frac{{b_2} }{ \sqrt{2}} -i\frac{ {b_3} }{ \sqrt{2}}\\
0 & \frac{1 }{ 4}+ \frac{b_4 }{ \sqrt{2}}-\frac{b_5 }{ 2} & 0 & 0 \\
0 & 0 & \frac{1 }{ 4}- \frac{b_4 }{ \sqrt{2}}-\frac{b_5 }{ 2}  & 0 \\
 \frac{{b_2} }{ \sqrt{2}}+i\frac{{b_3} }{ \sqrt{2}}    & 0 & 0 &
\frac{1 }{ 4}- \frac{b_1 }{ \sqrt{2}}+\frac{b_5 }{ 2}      \\
\end{array}
\right)~~~,
\label{eq:rorop1}
\end{eqnarray}
where the index $p$ stands for the direct product spin representation.
Diagonalizing the density operator ${\Op \rho_p}$ Cf.
Appendix B, leads to the eigenvalues of ${\Op \rho }$ which define the von Neumann probabilities:
\begin{eqnarray}
\begin{array}{c}
\lambda_1~ =~\frac{1}{4}-\frac{D }{ \sqrt{2}}+\frac{b_5}{2}\\
\lambda_2~=~\frac{1}{4}+\frac{b_4 }{ \sqrt{2}}-\frac{b_5}{2}\\
\lambda_3~=~\frac{1}{4}-\frac{b_4 }{ \sqrt{2}}-\frac{b_5}{2}\\
\lambda_4~=~\frac{1}{4}+\frac{D }{ \sqrt{2}}+\frac{b_5}{2}
\end{array}~~,
\label{eq:eigenvn}
\end{eqnarray}
where $D~=~\sqrt{b_1^2+b_2^2+b_3^2}$. Functionals of the density
operator such as entropy are calculated by the spectral theorem Cf. appendix A.

\section{Limit Cycles}
\label{sec:limit}

\begin{figure}[tb]
\vspace{0.66cm}
\hspace{3.cm}
\psfig{figure=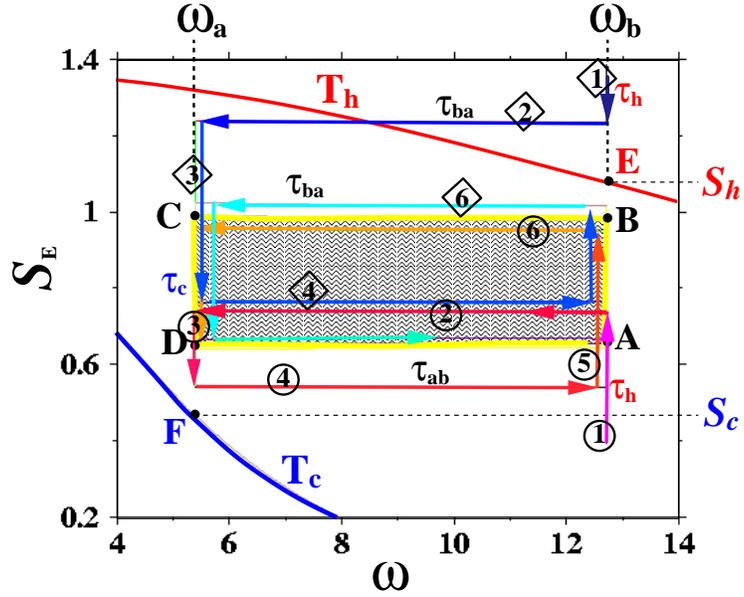,width=0.62\textwidth}
\vspace{0.5cm}
\caption{The limit cycle in the field entropy variables $(\omega ,{\cal S}_E )$.
Two initial states represented by points are indicated by 1 with the field value of
$\omega_b$. The sequence of branches starting from a cold initial temperature
are shown with circles around the numbers.
The working medium of the engine is put in contact with the hot bath
for a time duration of $\tau_h$ and heats up. For time $\tau_{ba}$
the frequency changes from $\omega_b$ to $\omega_a$ (branch 2).
Cooling due to contact with the cold bath is found on branch 3.
For clarity the arrows do not reach the values of $\omega_a$ and $\omega_b$.
The sequence of branches starting from a hot initial temperature
are shown with diamonds around the numbers.
The working medium of the engine is put in contact with the hot bath
for a time duration of $\tau_h$ and cools down to branch 2.
The contact with the cold branch further cools the engine (branch 3)
After going through approximately 3 cycles of the engine, the  two paths
appear to converge to the same limit cycle,
indicated by the ABCD rectangle.}
\label{fig:2}
\end{figure}

The heat engine's limit cycle is completely determined by the
external control parameters. This means that irrespective of the
initial state of the working medium, after running the engine
through many cycles of the control sequence, a limit cycle is
approached. This can be observed in Fig. \ref{fig:2} which
demonstrates that starting from two initial conditions, the engine
settles to the same limit cycle which is the fixed point at the bottom of the basin of attraction.

\subsection{Approach to the limit cycle}
\label{susec:approach}

\begin{figure}[tb]
\vspace{0.66cm} \hspace{2.cm}
\psfig{figure=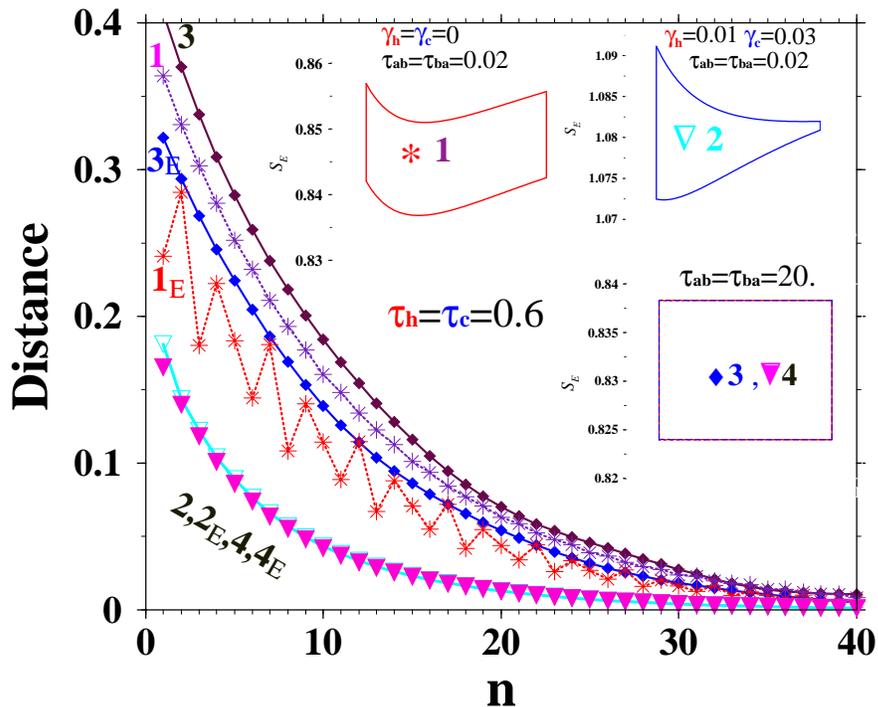,width=0.7\textwidth} \vspace{0.5cm}
\caption{The distance ${\cal D}(\Op \rho_n, \Op \rho_{lc})$ and
the energy distance ${\cal D}_E (\Op \rho_n, \Op \rho_{lc})$ from
the limit cycle as a function of accumulated cycles $n$. Four
example are shown numbered from 1 to 4. The corresponding inserts
show the limit cycle in the energy entropy ${\cal S}_E$ and
$\omega$ plane. The dashed lines indicated by $E$ correspond to
the energy distance ${\cal D}_E (\Op \rho_n, \Op \rho_{lc})$.
Cycles 1 and 3 are without dephasing. Cycle 2 and 4 include
dephasing on the {\em isochores} with $\gamma_h=0.01$ and
$\gamma_c=0.03$. All cycles spend the same time on the {\em
isochores} ($\tau_h= \tau_c=0.6$). The time on the {\em adiabats}
varies form very short, cycles 1 and 2, to very long, cycles 3 and
4.} \label{fig:3}
\end{figure}

The timescale of the approach of the engine to the limit cycle is
related to the number of accumulated cycles $n$, that are required
for the variables of the engine to approach their asymptotic
values. The different measures employed for this task are defined
in Sec. \ref{sec:sumquantherob}. The energy distance ${\cal D}_E
(\Op \rho_n, \Op \rho_{lc})$ Eq. (\ref{eq:fdistan}) and quantum
distance ${\cal D}(\Op \rho_n, \Op \rho_{lc})$ were used as
measures and are shown in Fig. \ref{fig:3}. The reference state
$\Op \rho_{lc}$ can be chosen on any point on the engine's
trajectory, on points {\bf  A ~B~ C ~D} for example or between
them. The same point on the trajectory is used to define the state
in the n'th iteration $\Op \rho_n$. It was found that the
distances are invariant to the choice of the chosen point on the
cycle's trajectory.

Explicitly the  quantum \em distance \rm, ${\cal D}(\Op \rho_n,\Op
\rho_{lc})$, Eq. (\ref{eq:quantdist2}) for the current working
medium becomes:
\begin{eqnarray}
{\cal D} (\Op \rho_n,\Op \rho_{lc}) =\sqrt{2 \left(1-(\sqrt{\zeta_1} + \sqrt{\lambda_2(n)
\lambda_3(lc)} +
 \sqrt{\lambda_3(n) \lambda_3(lc)} +  \sqrt{\zeta_4}) \right)}~~,
\label{eq:finaldist}
\end{eqnarray}
where the normalization is chosen to be $N=2$ and the eigenvalues
$\lambda$ are the  eigenvalues of the density operator defined in
Eq. (\ref{eq:eigenvn}) and,
\begin{eqnarray}
\zeta_{1,4}~=~Q \pm \sqrt{\left(\frac {Y D_n}{\sqrt{2}}
\right)^2+\left(\frac {x_n D_{lc}}{\sqrt{2}} \right)^2 +2x_n qY}~~,
\label{eq:zeteng1}
\end{eqnarray}
where $2 q$ is the scalar product between the $\{ \Op B \}$ components:
\begin{equation}
2 q= \left( \vec b_n \cdot \vec b_{lc} \right) = \sum_{j=1}^3 b_j(n) b_j(lc)
\label{eq:scalar}~~,
\end{equation}
and $x=\sqrt{\lambda_1 \lambda_4}$, $2 r=\frac{1}{2}+b_5$,
$y=2(r-x)/D$, $Y=r_{lc}+q y_n$ and finally $Q$ is the generalized
scalar product: $ Q=r_{n}\cdot r_{lc}+q$, ($D$ is defined by
following Eq. (\ref{eq:eigenvn})). Additional computational
details are given in Appendix \ref{ap:B}.

The distance to the limit cycle can also
be associated with the conditional entropy:
\begin{eqnarray}
\begin{array}{l}
 S(  \Op \rho_n | \Op \rho_{lc})= {\cal S}(\Op \rho_n)-\\
\left( \{ r_n - \frac{\sqrt{2} q} {D_{lc}} \} \log( \lambda_1 ({lc})) +
\lambda_2 (n) \log (\lambda_2 ({lc}))+  \lambda_3(n) \log (\lambda_3(lc)) +
 \{ r_n + \frac{ \sqrt{2} q} { D_{lc}} \} \log (\lambda_1(lc)) \right)~~,
\nonumber
\end{array}
\\
\label{eq:condenlcbase}
\end{eqnarray}
which becomes zero when $\Op \rho_n$ approaches the limit cycle.

In all cases studied the quantum distance ${\cal D}(\Op \rho_n, \Op \rho_{lc})$
monotonically approaches  zero with the increase in the
number $n$ of accumulated cycles,~ Cf. Fig. \ref{fig:3}. This was
found also to be true for the conditional entropy as predicted by
Eq. (\ref{eq:limit1}). The energy distance ${\cal D}_E (\Op
\rho_n, \Op \rho_{lc})$ as in case $1_E$ in Fig. \ref{fig:3}, shows
a non-monotonic periodic oscillations in the approach to the
limit cycle. With sufficient dephasing the density operator is
almost diagonal in the energy representation therefore the two
distances converge; ${\cal D}(\Op \rho_{n}, \Op \rho_{lc}  )={\cal
D}_E (\Op \rho_n, \Op \rho_{lc})$ Cf. cases 2 and 4 in Fig.
\ref{fig:3}.

Examining the measures of approach to the limit cycle Eq.
(\ref{eq:finaldist}) and (\ref{eq:condenlcbase}), it is found that
they have similar functional dependence on the expectation values of the set
of $\{\Op B\}$ operators. In particular the two functionals contain
the scalar product $q$. This indicates that all quantum convex
functionals of $\Op \rho$ will relax to the limit cycle in the
same rate.

\begin{figure}[tb]
\vspace{0.66cm} \hspace{2.cm}
\psfig{figure=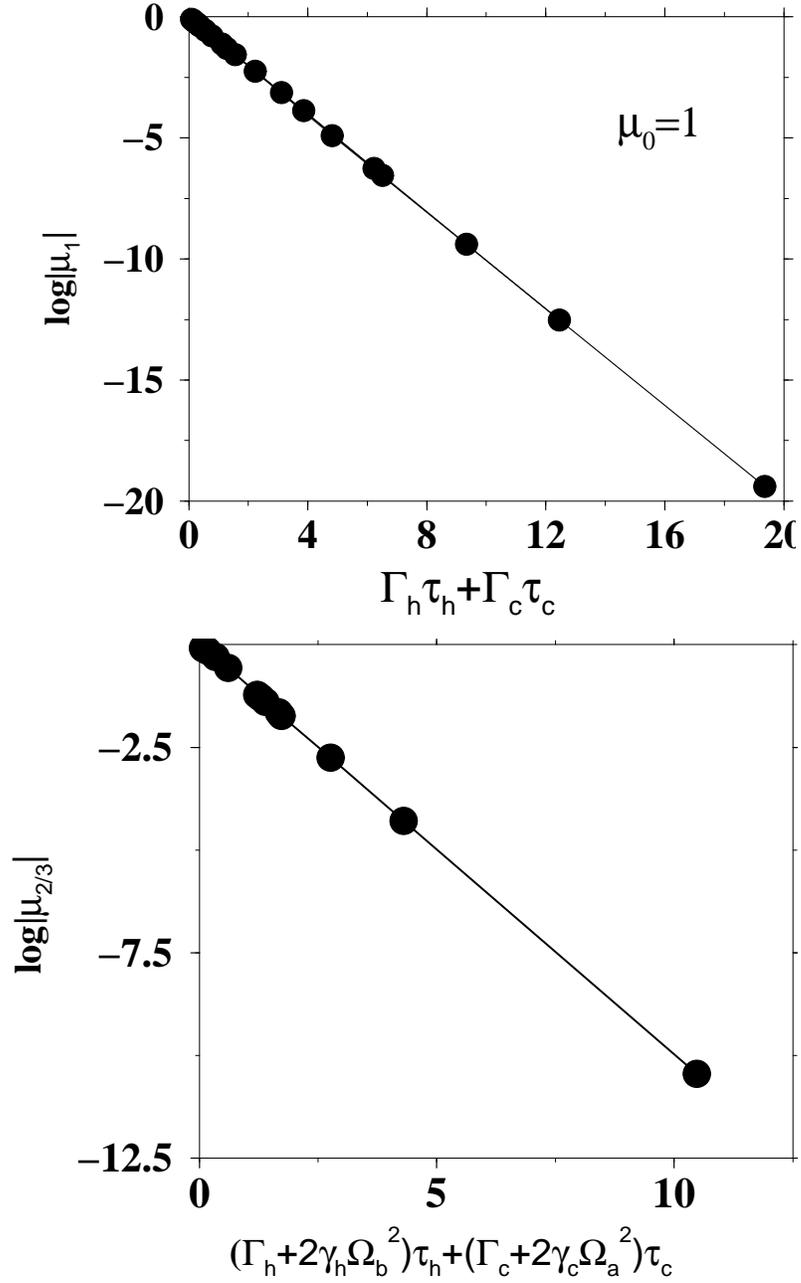,width=0.7\textwidth} \vspace{0.5cm}
\caption{Logarithms of the eigenvalues of the global
propagator  ${\cal U}_{cyc}$. Upper panel $\mu_1$ as a function of
the accumulated relaxation
$\Gamma_h \tau_{h}+\Gamma_c \tau_{c}$, the sum of the products of
the coupling to the heat/cold baths, $\Gamma_{h/c}$, with the time
allocation on the corresponding {\em isochore}, $\tau_{h/c}$. The
eigenvalue is independent of the time allocation on the
{\em adiabats}. Lower panel $\mu_{2/3}$ as a function of the
accumulated dephasing
$(\Gamma_h +2 \gamma_h \Omega_h^2 )\tau_{h}+(\Gamma_c +2 \gamma_c \Omega_c^2)
\tau_{c}$. The points on the graphs represent different choices of
parameters and time allocation on the branches.}
\label{fig:4}
\end{figure}

The dynamics of the set of operators $\{\Op B\}$ is determined by the
eigenvalues of the \em Global Operator \rm, ${\cal U}_{cyc}$,
 Eq. (\ref{eq:gprop}). The eigenvector
with the eigenvalue of $\mu_0=1$ represents the expectation values
of the limit cycle. The decay rate to the limit cycle depends on
the eigenvalues which are smaller than one. The eigenvalue $\mu_1
\le 1 $ was found to be real and as expected its eigenvector does
not include a component of the identity operator. Fig. \ref{fig:4}
shows the dependence of the eigenvalues $\mu$ of the \em Global
Operator \rm on the time spent on the {\em isochores} and the
coupling constants. The relationship is well fitted by $\mu_1 =
e^{-( \Gamma_h \tau_h+\Gamma_c \tau_c)}$. This means that a very
weak dependence on the dephasing rate $\gamma_{h/c}$ was found.
$\mu_1$ can then be interpreted as the relaxation rate per cycle
in the direction defined by the limit cycle vector. The
eigenvalues $\mu_{2/3} = |\mu_2| e^{\pm i\phi}$ are complex. Their
amplitude can be fitted to $|\mu_{2/3} |= e^{-\left( (\Gamma_h
+\gamma_h \Omega_h^2 )\tau_h+(\Gamma_c +\gamma_c \Omega_c^2)
\tau_c\right)}$. This suggests that $\mu_{2/3}$ represents the
rate of decay in a direction perpendicular to the direction of
limit cycle vector. The phase $\phi$ of $\mu_{2/3}$ is an
accumulated phase and was found to be linearly related to the time
allocated on the {\em adiabats}. The last two eigenvalues
associated with the block of $\Op B_4$ and $\Op B_5$ become
$\mu_4=\mu_1$ and $\mu_5=(\mu_1)^2$ ( Eq. (57) and (59) of
\cite{kosloff03} ).

The  analysis reveals that the rate of approach to the limit cycle
is determined by the accumulated dissipation on the the hot and
cold {\em isochores}. The eigenvalue $\mu_1$ plays the role of
the longitudinal relaxation analogue to $1/T_1$, while the
eigenvalues $\mu_{2/3}$ play the role of the transverse
relaxation or $1/T_2$.

\subsection{Properties of the limit cycle}

The fact that the limit cycle is closed imposes a strict periodic
constraint on all properties of the working medium. The
periodicity of the energy entropy and the von Neumann entropy are
a key to the understanding of the cycle performance.

\subsubsection{Minimal cycle time}

A limit cycle can only exist if the total internal entropy changes
on the four branches sums up to zero. What combinations of control
parameters such as the time allocations lead to a stable limit
cycle? This question can be addressed by searching for the
opposite conditions where a limit cycle cannot be closed. As was
described in Sec. \ref{subsec:mono} a limit cycle is obtained only
when there is a single invariant of the propagator ${\cal U}_{cyl}$. 
An extreme case is when no time is allocated to the hot
and cold {\em isochores }($\tau_c=\tau_h=0$). By construction the
map ${\cal U}_{cyl}$ is unitary and all eigenvalues are modulus 1.
Thus any initial state will oscillate indefinitely without
settling to a limit cycle. The next step in the investigation is
to allocate some time to the cold isochore $\tau_c \neq 0$ adding
a dissipative branch.
\begin{figure}[tb]
\vspace{0.66cm} \hspace{1.cm}
\psfig{figure=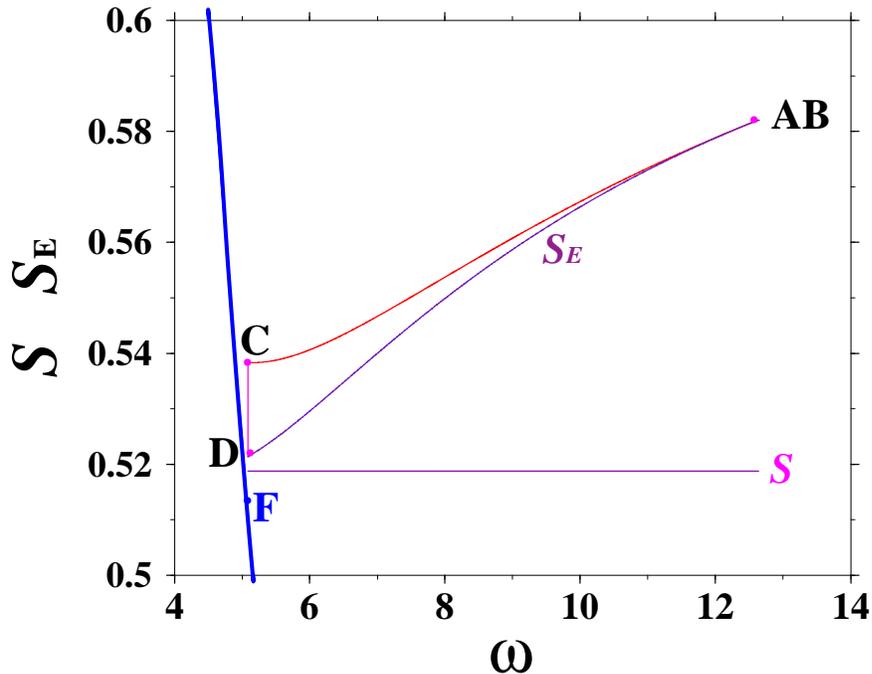,width=0.7\textwidth} \vspace{0.5cm}
\caption{The energy entropy ${\cal S}_E$ and the von Neumann
entropy ${\cal S}$ as a function of the field $\omega$ for a limit
cycle with time allocations on the {\em adiabats}
($\tau_{ba}=\tau_{ba}=0.03$) and the cold {\em isochore}
($\tau_c=0.6$). No time is allocated on the hot {\em isochore}.
Notice the horizontal line representing the von Neumann entropy.
Other parameters are: $J=2$, $T_h=7.5$, $T_c=1.5$,
$\omega_a=5.0836387$, $\omega_b=12.635485$, $\Gamma_c=1.7$. The
entropy production of the cycle is: $\Delta S^u=1.889 \cdot
10^{-2}$ and the power output is negative ${\cal P}=-4.293 \cdot
10^{-2}$.} \label{fig:6}
\end{figure}
Analyzing the eigenvalues of ${\cal U}_{cyl}$ shows the expected
invariant eigenvalue $\mu_0=1$. All other eigenvalues are smaller
than one meaning that a limit cycle exists. The entropy picture is
more surprising. The change in the von Neumann entropy ${\Delta S}_{ab/ba}$ 
on the two {\em adiabats} is zero (Cf. Fig. \ref{fig:6}) for
any point on the cycle trajectory. Therefore the only way the
cycle can be closed is that the  change of the von Neumann entropy
on the dissipative cold {\em isochore} is also zero. A
complementary picture is obtained by analyzing the energy entropy
${\cal S}_E$ along the trajectory. If the 
two {\em adiabats} are not an inverse of each other, 
then the evolution after a sequence of two {\em adiabats} (point D
$\rightarrow$ AB $\rightarrow$ C ) will lead to an increase in
${\cal S}_E$ (Cf. Fig. \ref{fig:6}). To close the cycle this
increase should be compensated by a decrease in the energy entropy
on the cold {\em isochore}. How do these seemingly contradictory
entropy balances coexist? The answer is hidden in the double role
the cold bath  plays in the entropy changes. If the state of the
working medium (point C in Fig. \ref{fig:6}) is hotter than the
temperature of the cold bath, heat will transfer from the working
medium to the cold bath , and thus decreasing the liquid's
entropy. On the other hand the contact with the bath forces
dephasing. This loss of phase increases the von Neumann entropy.
Therefore, there is a point on the cold {\em isochore} where the
decrease in the energy entropy is exactly compensated by the
entropy increase due to dephasing. This scenario defines the
stable limit cycle. This cycle cannot produce useful work. It
represents a device which converts work from the two {\em
adiabats} to heat dissipated in the cold bath in accordance with
the second law of thermodynamics.

\begin{figure}[tb]
\vspace{0.66cm} \hspace{1.cm}
\psfig{figure=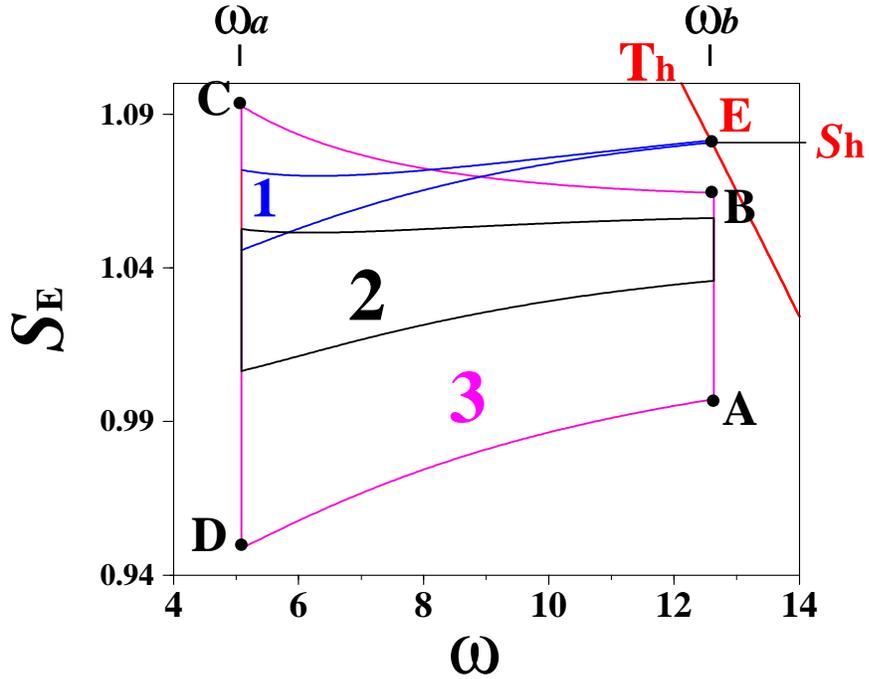,width=0.7\textwidth} \vspace{0.5cm}
\caption{Three cycles corresponding to the minimal cycle time "1",
zero power output "2"  and positive power output "3". The time
allocations on the \em isochores \rm are: on cycle "1"
$\tau_h=0.32$,~ $\tau_c=0.64$,~ on cycle "2": $\tau_h=0.581$,~
$\tau_c=1.1602$,~ and on cycle "3": $\tau_h=1.5$,~ $\tau_c=3.6$.
~The common parameter values on all three cycles are:
 $\tau_{ab}=0.05$,~ $\tau_{ba}=0.06$,~$J=2$,~$\Gamma_h=1.0048$,~
$\Gamma_c=0.10662$,~$\omega_b=12.63545$,~$\omega_a=5.0836387$,~$T_h=7.5,~
T_c=1.5$.}
\label{fig:5}
\end{figure}

Fig. \ref{fig:5} shows three cycles with time allocated to all
branches of the engine. The first cycle, "1" is an extension of
cycle of Fig. \ref{fig:6}. It has minimal contact with the hot
bath. The position of the limit cycle on the $(\omega, {\cal S}_E) $
plot is as far as possible from the cold equilibrium point. This
position maximizes the negative energy entropy change on the cold
{\em isochore}. Cycle "2" corresponds to a zero work cycle. The
work generated by extracting heat from the hot bath and ejecting
at the cold bath is balanced by the work against "friction". This
cycle defines the minimum operational cycle time. Cycle "3" is a
typical cycle with positive work output.

A search was carried out for time allocations where the cycle does
not  close, therefore no limit cycle exist. In general it was very
difficult to find such conditions. These atypical cases with
extremely small time allocations, were characterized by a non
uniqueness of the limit cycle. For slightly longer time
allocations a unique limit cycle was found which does not
represent an engine. The reason is that the cooling on the cold
{\em isochore} was not sufficient to dissipate the energy increase
on the {\em adiabats}. As a result additional cooling was required
on the hot {\em isochore}. The onset where heat is transfered by
the engine from the hot to the cold bath was termed in the
analogous engine based on a phenomenological description of
friction \cite{feldmann00}, as the minimal cycle time (Cf. Cycle "1"
in fig. \ref{fig:5}). Additional time allocation is required to 
reach the onset of positive work output (Cf. Cycle "2" in fig. \ref{fig:5}).

\subsubsection{Entropy Production}
\label{entpro123}

The position of the limit cycle is determined by the balance of
entropy. Since on the adiabatic branches the von Neumann entropy
is constant the increase in the entropy of the working medium
on the hot {\em isochore} should be exactly compensated on the
cold {\em isochore}.

Examining the entropy changes on the hot {\em isochore}, one can
compare the external entropy production $\Delta S^{ext}_{h}=-{\cal
Q}_h /T_h$ to the internal change. Schl\"ogl suggested
\cite{schlogl71,feldmann85}, that the internal entropy production
is related to the difference in the conditional entropy associated
with the equilibrium state:
\begin{equation}
\Delta S^{E}_h ~~=~~{\cal S}_E \left(\Op \rho_A | \Op
\rho_{eq}(T_h)\right) - {\cal S}_E \left(\Op \rho_B | \Op
\rho_{eq}(T_h)\right) ~~,
\label{eq:schlogl}
\end{equation}
where $\Op \rho_{A}$ is the state at point $A$ on the beginning of
the hot {\em isochore}, $\Op \rho_B$ is the state at point $B$ at
the end of the hot {\em isochore} and $\Op \rho_{eq}(T_h)$ is
the equilibrium state at point $E$.
Using the fact that the equilibrium density operator is diagonal in the
energy representation i.e. ${\cal S}_E \left(\Op \rho_A | \Op
\rho_{eq}(T_h)\right) = {\cal S}_E (\Op \rho_A )-E_A/T_h + \log Z$,
Eq. (\ref{eq:schlogl}) takes the form:
\begin{equation}
\Delta S^{E}_h ~~=~~{\cal S}_E \left(\Op \rho_A\right )
 - {\cal S}_E \left(\Op \rho_B \right) - \frac{{\cal Q}_h} {T_h}
\label{eq:schlogl-2}
\end{equation}
since ${\cal Q}_h=(E_b-E_a)$.

The full quantum characteristics representing the deviation
of the density operator from the energy representation
are obtained by the conditional entropy:
\begin{equation}
\Delta  S^u_h ~~=~~{\cal S} \left(\Op \rho_A | \Op
\rho_{eq}(T_h)\right) - {\cal S} \left(\Op \rho_B | \Op
\rho_{eq}(T_h)\right)~~=~~ {\cal S} \left(\Op \rho_A\right )
 - {\cal S} \left(\Op \rho_B \right) - \frac{{\cal Q}_h} {T_h}~~.
\label{eq:schlogl2}
\end{equation}
For very long cycle times and sufficient time allocation on the
{\em isochores } the system is diagonal in the energy
representation. As a result Eq. (\ref{eq:schlogl2}) and Eq.
(\ref{eq:schlogl}) are equivalent. The difference between the
internal entropy production $\Delta S^u_h $ and the external $\Delta
 S^{ext}_{h}=-{\cal Q}_h /T_h$ represents the entropy increase in
the working medium. The same measures can be applied to the cold
{\em isochore}. Summing the entropy changes on the two branches
leads to equality between the changes in external and internal
entropy production:
\begin{equation}
\Delta S^u_h+\Delta  S^u_c ~~=~~-\left(\frac{{\cal Q}_h}{ T_h} ~+~
\frac{{\cal Q}_c}{T_c} \right) ~~,\label{eq:entac}
\end{equation}
since the von Neumann entropy  ${\cal S}$ is constant on the {\em adiabats}. 
The energy entropy is not constant on the {\em adiabats}
leading to a different relation for the energy entropy production:
\begin{equation}
\Delta S^E_h+\Delta  S^E_c ~~=~~-\left(\frac{{\cal Q}_h}{ T_h} ~+~
\frac{{\cal Q}_c}{T_c} \right)+ \Delta {\cal S}^E_{ba}+\Delta
{\cal S}^E_{ab} ~~,\label{eq:entace}
\end{equation}
where $\Delta {\cal S}^E_{ab}$ is the change in energy entropy on
the {\em adiabat} which can be interpreted as the entropy
generation on the {\em adiabats}.

\section{Conclusions}
\label{sec:discussion}

Summarizing the study is best carried out by addressing the
questions raised in the introduction.

\subsection{How do the control parameters characterize the approach to the limit cycle?}

The existence of a limit cycle is subject to there being a unique
invariant of the global propagator. The invariant has an
eigenvalue $\mu_0=1$ and its eigenvector is expressed via the
expectation values of $\Op B_1$,  $\Op B_2$  $\Op B_3$ and $\Op I$
in the limit cycle.

Quantum measures were developed to characterize the approach to
the limit cycle: the conditional entropy and the quantum distance.
These measures show a monotonic approach to the limit cycle. The
projected measures such as the energy distance or energy
conditional entropy can show an oscillatory approach to the limit
cycle. Close to the limit cycle the rate of approach of all
measures converge to the same value. The quantum distance was
always larger than the probability distance associated with the
measurement of energy. Dephasing eroded the deviation between
the two distances.

Longitudinal and a transverse modes of approach to the limit cycle
could be identified. The rate of approach is associated to the
eigenvalues of the propagator. The eigenvalue $\mu_1$ determines
the longitudinal relaxation rate. $\mu_1$ exponentially depends on
the accumulated energy relaxation on the hot and cold {\em
isochores} $\mu_1 \propto e^{-(\Gamma_h \tau_{h}+\Gamma_c
\tau_{c})}$. The transverse rate of approach is associated with
the eigenvalues $\mu_{2/3}$. Their magnitude depends on the
accumulated dephasing on the hot and cold {\em isochores}
$|\mu_{2/3}| \propto e^{-\left( (\Gamma_h +\gamma_h \Omega_h^2
)\tau_h+ (\Gamma_c +\gamma_c \Omega_c^2) \tau_c\right)}$. The
phase $\phi$ of $\mu_{2/3}$ is linear in the time allocation
$\tau_{ab}+\tau_{ba}$. The dependence of the rate of relaxation on
other parameters such as $J$ was found to be weak.

\subsection{Can conditions be found for non-existence of the limit cycle?}

When no time is allocated to the hot and cold {\em isochores}
$\tau_h=\tau_c=0$ then the evolution is unitary and the modulus of
all eigenvalues of the propagator become 1. As a result no unique
limit cycle can be found. For very short times allocated to the
{\em isochores}, two eigenvalues of the cycle propagator became
equal to one, again no limit cycle is obtained in these
conditions. Therefore there exist a $\delta \tau \ll 1$ range of
time allocation on the {\em isochores} for which no unique limit
cycle can be closed.

\subsection{What are the irreversible properties of the limit cycle?}

Heat transport between the working medium and the baths is a
common source of irreversibility for all realistic heat engines.
If this is the only source of entropy generation the engine is
classified as endoreversible \cite{curzon75,salamon01}, meaning that
entropy is only generated on the interface and that the internal
operation is reversible. The dissipative forces accompanying heat
transfer were found to be sufficient to drive a quantum two-level
endoreversible heat engine to a limit cycle \cite{feldmann96}.

Friction is an additional source of irreversibility for all
realistic heat engines, characterized by an internal entropy
production. The heat generated by friction eventually has to be
disposed in the cold bath. The performance of the present first
principle quantum engine has been shown to be limited by a
friction like phenomena \cite{feldmann00}. The key to the
understanding of the quantum origin of friction lies in the
difference between the energy entropy ${\cal S}_E$ and the von
Neumann entropy ${\cal S}$. Since the von Neumann entropy is
constant on the {\em adiabats} one could classify the model as
endoreversible  (Cf. Eq. (\ref{eq:entac})). Following the engines cycle by observing its
energy changes shows characteristics of entropy generation on the
{\em adiabats} (Cf. Eq. (\ref{eq:entace})).

The most illuminating case which characterizes the irreversible
character due to the nonadiabatic dynamics is a cycle composed of
two {\em adiabats} and only a cold {\em isochore} as displayed in
Fig. \ref{fig:6}. External work is converted to internal heat
which is dissipated to the cold bath. The only phenomena that fits
this behavior is friction. Surprisingly, the von Neumann entropy
for the complete cycle trajectory is constant. This is in contrast
to a power producing cycle where the von Neumann entropy changes
on the {\em isochores}.   A detailed analysis of the von Neumann
entropy change on the cold {\em isochore} performed in the energy
representation unravel the picture. A decrease in the entropy of
diagonal elements, equivalent to $\Delta {\cal S}_E$ due to
cooling of the working medium is exactly compensated  by an
entropy increase due to dephasing i.e. loss of the nondiagonal
elements. It seems therefore that friction is the result of the
interplay between the unitary evolution on the {\em adiabats } and
the dissipative dynamics on the cold {\em isochore}. Friction is
found only when the state of the quantum engine deviates from a
diagonal energy representation. Such dynamics are a consequence of
the nonadiabatic operation conditions caused by the
noncommutability of the working medium Hamiltonian at different
points  along the cycle trajectory. These observations are the
basis for a quantum control of friction which will be presented in
a future study.

\begin{acknowledgments}
We thank Lajos Di{\'o}si and Jeffrey Gordon many discussions and a
critical reading. This work was supported by the Israel Science
Foundation.
\end{acknowledgments}

\appendix

\section{Analytical solution of the propagator on the \em adiabats \rm}
\label{ap:ansolprop}

The analytic solution for the propagator on the {\em adiabats}
is based on the Lie group
structure of the $\{\bf \hat B \}$ operators. The unitary evolution
operator $\bf \hat U (t)$ for an explicitly time dependent
Hamiltonians is
obtained from the Schr\"odinger equation:
\begin{eqnarray}
-i\frac{d}{dt}
{{\bf \hat  U} (t)}  ~=~
{\bf \hat H} (t){\bf \hat U} (t),   ~~~~~~~~
 {\bf \hat U} (0)={\bf \hat I }~~~.
\label{weinorde}
\end{eqnarray}
The propagated set of operators becomes:
\begin{eqnarray}
{\vec {\bf \hat B}} (t)    ~=~  {\bf \hat U} (t)
\vec {\bf \hat B} (0) {\bf \hat U}^{\dagger} (t)~=~
  {\cal U}_a(t)    {\vec {\bf \hat B}}  (0) ~~~,
\label{sandwich}
\end{eqnarray}
and is related to the super-evolution operator ${\cal U}_a(t)$.
Based on the group structure Wei and Norman, \cite{weinorman63}
constructed a solution to Eq. (\ref{weinorde}) for any
operator ${\bf \hat H}$ which can be written as a linear combination
of the operators in the closed Lie algebra
${\bf \hat H}  (t)  ~=~ \sum_{j=1}^m h_j(t){\bf \hat B}_i  $,
where the $h_i(t)$ are scalar functions of $t$.
In such a case the unitary evolution operator  ${\bf \hat U} (t)$
can be represented in the product form:
\begin{eqnarray}
{\bf \hat U} (t)  ~=~
\prod_{k=1}^{m} \exp(\alpha_k(t){\bf \hat B} _k) ~~~.
\label{weinorde2}
\end{eqnarray}
The product form Eq. (\ref{weinorde2}) substitutes
the time dependent operator equation (\ref{weinorde})
with a set of scalar differential
equations for the functions $\alpha_k(t)$.
Writing the unitary evolution operator explicitly leads to:
\begin{eqnarray}
{\bf \hat U} (t)   ~=~
 \exp(i\frac{ \alpha_1(t) }{ \sqrt{2} } {\bf \hat B_1} )
\exp(i \frac{\alpha_2(t)
}{ \sqrt{2}}{\bf \hat B_2} )
\exp(i \frac{\alpha_3(t) }{\sqrt{2}}{\bf \hat B_3} )
\label{weinorde3}
\end{eqnarray}
The $\sqrt{2}$ factor is introduced for technical reasons.
Based on the group structure \cite{weinorman63}
Eq. (\ref{weinorde}) leads to the following
set of differential equations for the coefficients $\alpha$:
\begin{eqnarray}
\dot \alpha_1= \sqrt{2} \omega(t)+ \sqrt{2} J (\frac{\sin(\alpha_1)
\sin(\alpha_2) }{ \cos( \alpha_2)})~;~~
\dot \alpha_2=  \sqrt{2} J \cos(\alpha_1)~;~~
\dot \alpha_3= \frac{ \sqrt{2} J \sin(\alpha_1)}{ \cos(\alpha_2) }~~~.
\label{mateq3}
\end{eqnarray}

\section{The Density Operators}
\label{subsec:densitop}

\subsection{Functions of the Density Operators}
\label{subsec:funcdenop}

\subsubsection{Computation of $\Op \rho_p^{\frac{1}{2}}$ and $ \log \Op \rho_p$}

First $\Op \rho_p$, Eq. (\ref{eq:rorop1}), is diagonalized by the
unitary matrices ${Q_p}~,~ Q^{\dagger}_p$:
\begin{eqnarray}
\begin{array}{c}
{ Q_p}
\end{array}~~=~\left(
\begin{array}{cccc}
-\frac{(b_2+ib_3) }{\sqrt{2D(D+b_1)}}    &0 & 0 &
\sqrt{\frac{(D+b_1)}{ 2 D}}  \\
 0 &1 & 0 & 0 \\
 0 &0 & 1 & 0  \\
\frac{ (b_2+ib_3) }{\sqrt{ 2 D(D-b_1)}}  & 0 & 0 &
\sqrt{\frac{(D-b_1) }{ 2 D}}  \\
\end{array}
\right)~~,
\label{LEFTQRHO}
\end{eqnarray}
leading to:
\begin{eqnarray}
Q_p \Op \rho_p Q^{\dagger}~=~
\left(
\begin{array}{cccc}
\frac{1}{4}-\frac{D }{ \sqrt{2}}+\frac{b_5}{2}&0&0&0 \\
0&\frac{1}{4}+\frac{b_4 }{ \sqrt{2}}-\frac{b_5}{2}&0&0 \\
0&0&\frac{1}{4}-\frac{b_4 }{ \sqrt{2}}-\frac{b_5}{2}&0 \\
0 & 0 &0 & \frac{1}{4}+\frac{D }{ \sqrt{2}}+\frac{b_5}{2}   \\
\end{array} \right)~=~\Op \rho_{vn}
\label{DIAGrho}
\end{eqnarray}
where $D~=~\sqrt{b_1^2+b_2^2+b_3^2}$, and  $\lambda_i$ are the
eigenvalues of $\Op \rho$ which are the von Neumann probabilities,
Cf.  Eq. (\ref{eq:eigenvn}).

The eigenvalues of ${\Op \rho^{\frac{1}{2}}}$ become
$\lambda_1^{\frac{1}{2}}$,
$\lambda_2^{\frac{1}{2}}$, $\lambda_3^{\frac{1}{2}}$,
$\lambda_4^{\frac{1}{2}}$. From Eq. \ref{DIAGrho} one has:~~ $\Op
\rho_p~=~Q^{\dagger}_p \Op \rho_{vn} Q_p$,~~  therefore $\Op
\rho_p^{\frac{1}{2}}~=~Q^{\dagger}_p \Op \rho_{vn}^{\frac{1}{2}}
Q_p$, and $\log \Op \rho_p~=~Q^{\dagger}_p \log \Op \rho_{vn}
Q_p$. Explicitly:
\begin{eqnarray}
\begin{array}{c}
{ \Op \rho_p^{\frac{1}{2}}}
\end{array}~~=~~
\left(
\begin{array}{cccc}
\frac{\lambda_4^{\frac{1}{2}}+\lambda_1^{\frac{1}{2}}}{2}+
\frac{b_1(\lambda_4^{\frac{1}{2}}-\lambda_1^{\frac{1}{2}})}{2D}
&0&0&\frac{(b_2-ib_3)(\lambda_4^{\frac{1}{2}}-
\lambda_1^{\frac{1}{2}})}{2D}\\
0&\lambda_2^{\frac{1}{2}}&0&0 \\
0&0&\lambda_3^{\frac{1}{2}} &0 \\
\frac{(b_2+ib_3)(\lambda_4^{\frac{1}{2}}-\lambda_1^{\frac{1}{2}})}{2D}
& 0 &0 & \frac{\lambda_4^{\frac{1}{2}}+\lambda_1^{\frac{1}{2}}}{2}-
\frac{b_1(\lambda_4^{\frac{1}{2}}-\lambda_1^{\frac{1}{2}})}{2D}  \\
\end{array} \right)
\label{rhophalf}
\end{eqnarray}
and
\begin{eqnarray}
\begin{array}{c}
{\log \Op \rho_p}
\end{array}~~=~~
\left(
\begin{array}{cccc}
\frac{\log \lambda_4+ \log \lambda_1}{2}+
\frac{b_1(\log \lambda_4-\log \lambda_1)}{2D}
&0&0&\frac{(b_2-ib_3)(\log \lambda_4-
\log \lambda_1)}{2D}\\
0&\log \lambda_2 &0&0 \\
0&0&\log \lambda_3 &0 \\
\frac{(b_2+ib_3)(\log \lambda_4-\log \lambda_1)}{2D}
& 0 &0 & \frac{\log \lambda_4+\log \lambda_1}{2}-
\frac{b_1(\log \lambda_4-\log \lambda_1)}{2D}  \\
\end{array} \right)
\label{logrho}
\end{eqnarray}

\subsubsection{Computation of $\Op \rho_e^{\frac{1}{2}}$ and $ \log \Op \rho_e$}

To get $\Op \rho$ in the energy picture $\Op \rho_p$ is
transformed  by the matrix ${\cal C}$ which diagonalized the
Hamiltonian, see \cite{kosloff03}. Denoting
$\Omega=\sqrt{\omega^2+J^2}$,~ $\mu=\sqrt{\frac{\Omega - \omega }{
2 \Omega}}$,~ and $\chi=\sqrt{\frac{\Omega + \omega }{ 2
\Omega}}$,~~~ $\cal C$ becomes \cite{kosloff03}:
\begin{eqnarray}
\begin{array}{c}
{\cal C}
\end{array}~~=~\left(
\begin{array}{cccc}
-\mu  &0 & 0 & \chi \\
 0 &1 & 0 & 0 \\
 0 &0 & 1 & 0  \\
\chi  & 0 & 0 & \mu  \\
\end{array}
\right)
\label{EIGHA}
\end{eqnarray}
Observing, that ${\cal C}{\cal C}~=~ I$, leads to:
${\Op \rho_e}={\cal C} {\Op \rho_p} {\cal C} $,
\begin{eqnarray}
\begin{array}{c}
\Op \rho_{e}
\end{array} ~~=
~~~  \left(
\begin{array}{cccc}
\frac{1 }{ 4} - \frac{ E }{ \Omega \sqrt{2} }+\frac{b_5 }{ 2} & 0 & 0 &
+ \frac{ib_3 }{ \sqrt{2}}
-\frac{ J b_1 }{ \Omega \sqrt{2} }+
\frac{ \omega b_2 }{ \Omega \sqrt{2} }        \\
0 & \frac{1 }{ 4}+ \frac{b_4 }{ \sqrt{2}}-\frac{b_5 }{ 2} & 0 & 0 \\
0 & 0 & \frac{1 }{ 4}- \frac{b_4 }{ \sqrt{2}}-\frac{b_5 }{ 2}  & 0 \\
 -\frac{i b_3}{ \sqrt{2}}
-\frac{J b_1 }{ \Omega \sqrt{2} }+
\frac{ \omega b_2 }{ \Omega \sqrt{2}}  & 0 & 0 &
\frac{1 }{ 4}+ \frac{ E }{ \Omega \sqrt{2}} +\frac{b_5 }{ 2}      \\
\end{array}
\right)~~~,
\label{rorop}
\end{eqnarray}
where $E= \omega b_1+Jb_2$. In equilibrium, the off-diagonal elements vanish.

${\Op \rho_p}={\cal C} {\Op \rho_e} {\cal C} $  , therefore~~:
${\Op \rho_{vn}}=Q_p {\Op \rho_p}Q^{\dagger}_p= Q_p{\cal C} {\Op
\rho_e} {\cal C} Q^{\dagger}_p$. It follows, that the
diagonalizing matrices of $\Op \rho_e$, become:~ $ Q_e=~Q_p{\cal
C}$, ~and ~ $Q^{\dagger}_e~=~{\cal C} Q^{\dagger}_p$. As a
result~~ $ \Op \rho_e^{\frac{1}{2}}=Q^{\dagger}_e {\Op
\rho_{vn}^{\frac{1}{2}}} Q_e $ ~~and~~~
$\log({\rho_e})=Q^{\dagger}_e \log( {\Op \rho_{vn}}) Q_e$.
Explicitly:
\begin{eqnarray}
\begin{array}{c}
{\Op \rho_e^{\frac{1}{2}}}
\end{array} ~~=
~~~  \left(
\begin{array}{cccc}
\frac{\lambda_4^{\frac{1}{2}}+\lambda_1^{\frac{1}{2}}
 }{2 }-\frac{E(\lambda_4^{\frac{1}{2}}-\lambda_1^{\frac{1}{2}})}
{2D \Omega}  & 0 & 0 &-\frac{(\lambda_4^{\frac{1}{2}}-\lambda_1^
{\frac{1}{2}})\left(\omega b_2-Jb_1+i \Omega b_3 \right)}
{2D \Omega }  \\
0 & \lambda_2^{\frac{1}{2}} & 0 & 0 \\
0 & 0 & \lambda_3^{\frac{1}{2}} & 0 \\
-\frac{(\lambda_4^{\frac{1}{2}}-\lambda_1^
{\frac{1}{2}}) \left(\omega b_2-Jb_1-i \Omega b_3 \right)}
{2D \Omega }                     & 0 & 0 &
\frac{\lambda_4^{\frac{1}{2}}+\lambda_1^{\frac{1}{2}}
 }{2 }+\frac{E(\lambda_4^{\frac{1}{2}}-\lambda_1^{\frac{1}{2}})}
{2D \Omega} \\
\end{array}
\right)~~~.
\label{rhoehalf}
\end{eqnarray}
and
\begin{eqnarray}
\begin{array}{c}
{\log \Op \rho_e}
\end{array} ~~=
~~~  \left(
\begin{array}{cccc}
\frac{\log \lambda_4+\log \lambda_1
 }{2 }-\frac{E(\log \lambda_4-\log \lambda_1)}
{2D \Omega}  & 0 & 0 &-\frac{(\log \lambda_4-\log \lambda_1)
\left(\omega b_2-Jb_1+i \Omega b_3 \right)}
{2D \Omega }  \\
0 & \log \lambda_2 & 0 & 0 \\
0 & 0 & \log \lambda_3 & 0 \\
-\frac{(\log \lambda_4-\log \lambda_1)
\left(\omega b_2-Jb_1-i \Omega b_3 \right)}
{2D \Omega }                     & 0 & 0 &
\frac{\log \lambda_4+\log \lambda_1
 }{2 }+\frac{E(\log \lambda_4-\log \lambda_1)}
{2D \Omega} \\
\end{array}
\right)~~~.
\label{lnrop}
\end{eqnarray}
Any function of the density matrix can be computed  by the diagonalizing vectors of the density matrix.

\subsection{Additional details of quantum distance}
\label{ap:B}

In subsection \ref{susec:approach}, a closed form expression
was obtained for the quantum distance, Eq. (\ref{eq:finaldist}).
For the computation the polarization frame, $\Op \rho_p$, was used.

The operator
${\Op M}=(\Op \rho)^{ \frac{1}{2}} \Op \rho_{ref} (\Op \rho)^{ \frac{1}{2}}$
required in Eq. (\ref{eq:quantdist2}) is first computed.
\begin{eqnarray}
\nonumber
\begin{array}{c}
{\Op M}
\end{array}~~=~~
\end{eqnarray}
\begin{eqnarray}\left(
\begin{array}{cccc}
Q+\frac{b_1(n)}{\sqrt{2}}(Y+x_n)&0&0&(\frac {b_2(n)}{\sqrt{2}}-i\frac{b_3(n)}{\sqrt{2}})Y+(\frac{b_2(lc)}{\sqrt{2}}-i\frac {b_3(lc)}{\sqrt{2}})x_n\\
0&\Lambda_2 &0&0\\
0&0&\Lambda_3&0\\
(\frac{b_2(n)}{\sqrt{2}}+i\frac {b_3(n)}{\sqrt{2}})Y+(\frac{b_2(lc)}{\sqrt{2}}+i\frac {b_3(lc)}{\sqrt{2}})x_n&0&0&Q-\frac{b_1(n)}{\sqrt{2}}(Y+x_n)\\
\end{array}\right)
\label{mdis1}
\end{eqnarray}
where the notations of Eqs. (\ref{eq:finaldist}) was used,
(\ref{eq:zeteng1})~~and (\ref{eq:scalar}), and where\\
$\Lambda_2=\lambda_2^{\frac{1}{2}}(n)\lambda_2^{\frac{1}{2}}(lc)$,~and~
$\Lambda_3=\lambda_3^{\frac{1}{2}}(n)\lambda_3^{\frac{1}{2}}(lc)$.~

Calculation of  ${\cal D} (\Op \rho_n,\Op \rho_{lc})$,~ requires the  value of $tr \{ \sqrt{\Op M} \}$.
The matrix representation of ${\Op M}$ breaks up into two internal and external 2x2 sub-matrices.
Denoting the eigenvalues of  the external ${\Op M}_{1,4}$ submatrix by $\zeta_i$, one has:
\begin{eqnarray}
\zeta_{1,4}~=~Q \pm \sqrt{\left(\frac {YD_n}{\sqrt{2}} \right)^2+\left(\frac {x_n D_{lc}}{\sqrt{2}} \right)^2
+2x_nqY}
\label{zeteng1}
\end{eqnarray}

\pagebreak

\end{document}